\documentclass[aps,prl,twocolumn,amsmath,amssymb,superscriptaddress]{revtex4-1}

\usepackage{blindtext}
\usepackage{xfrac}
\usepackage[percent]{overpic}
\usepackage{graphicx}
\usepackage{amssymb}
\usepackage{color}
\usepackage{psfrag}
\usepackage{ifsym}
\usepackage{epstopdf}
\usepackage{soul}
\usepackage{tabularx}
\usepackage{hyperref}
\usepackage{amsmath,amssymb}
\usepackage{float}
\usepackage{dcolumn}

\newcolumntype{u}[1]{D{!}{\,\pm\,}{#1} }

\newcommand{\bra}[1] {\langle #1 |}
\newcommand{\ket}[1] {| #1 \rangle}
\newcommand{\Gt}{$G^{(2)}_{s,i}(\tau)\,$}

\newcommand{\degr}{$^{\circ}$}
\newcommand{\myhref}[3][black]{\href{#2}{\color{#1}{#3}}}

\begin{document}
\title{Sub-Megahertz Single Photon Source}

\author{Markus Rambach}\email{m.rambach@uq.edu.au}
\affiliation{ARC Centre for Quantum Computation \& Communication Technology, and ARC Centre for Engineered Quantum Systems, and School of Mathematics and Physics, University of Queensland, 4072 Brisbane, QLD, Australia.}
\author{Aleksandrina Nikolova}
\affiliation{ARC Centre for Quantum Computation \& Communication Technology, and ARC Centre for Engineered Quantum Systems, and School of Mathematics and Physics, University of Queensland, 4072 Brisbane, QLD, Australia.}
\author{Till J. Weinhold}
\affiliation{ARC Centre for Quantum Computation \& Communication Technology, and ARC Centre for Engineered Quantum Systems, and School of Mathematics and Physics, University of Queensland, 4072 Brisbane, QLD, Australia.}
\author{Andrew G. White}
\affiliation{ARC Centre for Quantum Computation \& Communication Technology, and ARC Centre for Engineered Quantum Systems, and School of Mathematics and Physics, University of Queensland, 4072 Brisbane, QLD, Australia.}


\begin{abstract}

We report $100$\% duty cycle generation of sub-MHz single photon pairs at the Rubidium D$_1$ line using cavity-enhanced spontaneous parametric downconversion. The temporal intensity cross-correlation function exhibits a bandwidth of $666 \pm 16$~kHz for the single photons, an order of magnitude below the natural linewidth of the target transition. A half-wave plate inside our cavity helps to achieve triple resonance between pump, signal and idler photon, reducing the bandwidth and simplifying the locking scheme. Additionally, stabilisation of the cavity to the pump frequency enables the $100$\% duty cycle. These photons are well-suited for storage in quantum memory schemes with sub-natural linewidths, such as gradient echo memories.

\end{abstract}


\maketitle

\section{Introduction}

Quantum technologies are primed to revolutionise information processing, with large companies already investing in basic quantum computing devices. While these potentially offer local computational enhancement, distributing quantum information will also require dedicated quantum networks designed to faithfully transmit quantum bits (qubit) \cite{Kimble2008}. Photons are the natural choice for information carriers due to their high mobility and low interaction with the environment. Overcoming the inevitable transmission losses in such networks demands repeater nodes with  entanglement swapping operations \cite{Duan2001,Simon2007,Choi2008,Munro2010}. The absence of high fidelity entangling gates in linear optical quantum computing (LOQC), and true on-demand sources means that the repeater node will also require noiseless amplification as well as a memory to hold the photonic qubit. As photons are notoriously difficult to store locally, the transfer of the qubit from the photonic state to a locally storable qubit, such as atoms \cite{Philips2001,Hosseini2011a}, ions \cite{Kielpinski2001}, Nitrogen-vacancy centres in diamonds \cite{Heshami2014} or rare-earth doped solids \cite{Afzelius2009,Gundogan2015}, is essential. For this kind of hybridisation the spectral properties of the photons and the chosen transition of the storage qubit need to be matched. As the spectral properties of the memories are much more limited in their tunability, the single photon source needs to be engineered to match the wavelength and the bandwidth of the atomic transition. 

\begin{figure*}[!t]
\centering
\fbox{\includegraphics[width=0.7\linewidth]{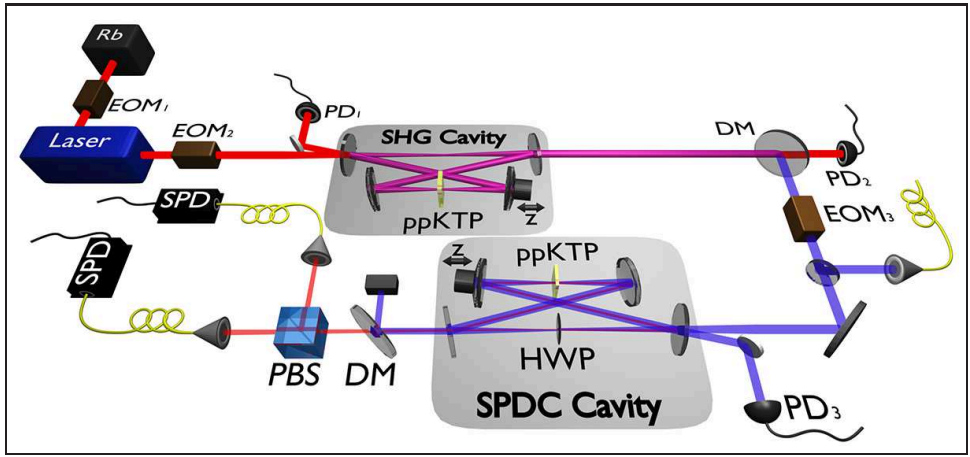}}
\caption{ Schematic experimental setup. The laser is stabilised to the SHG cavity which itself is stabilised to the Rb  D$_1$ transition for absolute frequency stability. The generated light at $397.5$~nm is separated from $795$~nm light at a DM and pumping the type-II SPDC cavity. Finally the single photons are split on a PBS and fibre-coupled for further processing. Rb, Rubidium spectroscopy cell; EOM$_{1,2,3}$, electro-optic modulators; PD$_{1,2,3}$, photodiodes; SHG, second-harmonic generation; ppKTP, non-linear crystal; $\stackrel{\text{z}}{\Longleftrightarrow}$, piezo-electric transducer; DM, dichroic mirror; SPDC, spontaneous parametric downconversion; HWP, half-wave plate; PBS, polarising beamsplitter; SPD, single photon detectors. }
\label{fig:setup}
\end{figure*}

Suitable photons can be generated through four-wave mixing in a magneto-optical trap \cite{Zhao2014, Liao2014}; however, this requires substantial experimental effort, and the duty cycles are low. Alternatively, the long-standing gold standard for single photon generation, spontaneous parametric downconversion (SPDC) \cite{Kwiat1995}, offers high-purity heralded single photon generation, as well as flexible emission wavelengths. However, the spontaneous nature of the downconversion process, combined with energy conservation and phase matching conditions results in a frequency spectrum typically on the order of 100s of GHz up to THz, several orders of magnitude larger than the bandwidths of atomic transitions. This can be compensated either by spectral filtering \cite{Haase2009,Clausen2011,Luo2015}, which severely reduces the available brightness, or by using an optical cavity to enhance the probability of creating the photons in the spectral and spatial resonator mode \cite{Ou1999,Wolfgramm2008,Bao2008,Scholz2009,Zhang2011,Pomarico2012,Fekete2013,Zhou2014}.

One of the most promising candidates for quantum memories to date is the gradient echo memory (GEM) \cite{Hosseini2011,Hosseini2012,Sparkes2010,Sparkes2013}, with recall fidelities of up to $98$\% \cite{Hosseini2011a} and coherence times of up to $195\ \mu$s \cite{Sparkes2013} in Rubidium, and even longer storage times in Rare Earth ion crystals \cite{Zhong2015}. To achieve these high fidelities and storage times, the spectral bandwidth of the photons needs to be well below the natural linewidth, e.g., sub-MHz for Rubidium. Previous cavity-based SPDC sources have achieved bandwidths comparable to atomic linewidths, but divide their operation time into stabilisation and photon production phases, resulting in typical duty cycles $< 50$\%. Fekete et al. reported the so far narrowest photons from SPDC---around $2$~MHz \cite{Fekete2013}, still unsuitable for efficient GEM---and only one source has demonstrated $100$\% duty cycle \cite{Scholz2009}.

Here we report on a triply resonant sub-megahertz source of orthogonally polarised single photon pairs with a $100$\% duty cycle. The photons are created via cavity-enhanced SPDC at the $795$~nm Rb wavelength, with bandwidths for signal and idler of $666 \pm 16$~and $667 \pm 15$~kHz, both matching the D$_1$ line in Rb used in GEM. 


\section{Experimental setup}

Generating photons resonant to atomic transitions requires the length of the SPDC cavity to be locked to an absolute frequency reference. Previous sources commonly achieved this during the feedback part of a cycle by stabilising the cavity to a laser beam on resonance with the atomic ensemble. As the reference beam is at the target frequency, no single photons can be observed during this phase. For the remainder of the cycle the cavity is not actively stabilised and only the pump light is present, generating resonant single photon pairs. This procedure reduces the duty cycle to well below $100$\%. Our system overcomes this issue by stabilising the SPDC cavity to the pump light, which itself is referenced back to the targeted Rb transition, as explained in detail below.

The schematic experimental setup is shown in Fig.~\ref{fig:setup}. The starting point is a continous-wave amplified diode laser (Toptica TA Pro) emitting laser light at $795$~nm (subscript $r$ for red) through two output ports. The light from the master port is modulated by a $1.25$~MHz RF signal with small modulation depth in an electro-optic modulator (EOM$_1$), and enters a Rubidium cell (TEM CoSy), where Doppler-free spectroscopy is performed. The output of the amplified slave port is modulated at $12.5$~MHz  (EOM$_2$) and coupled into the cavity for second-harmonic generation (SHG). The cavity enhances the efficiency of the frequency doubling inside a $20$~mm long periodically poled potassium titanyl phosphate (ppKTP) crystal to $38$\%, and the created laser light at $397.5$~nm (subscript $b$ for blue) is subsequently pumping the downconversion. The free spectral range of the cavity is $FSR_{r,SHG} \approx 278$~MHz with a finesse of ${\cal F}_{r,SHG} \approx 100$ (for detailed numbers including uncertainties see supplementary material). The laser system is stabilised to the error signal, derived from the SHG cavity reflection on PD$_1$ (home-made fast photodiode), using the Pound-Drever-Hall (PDH) technique \cite{Drever1983,Black2001} which is utilised for all frequency locks in the experiment. The feedback loop simultaneously adjusts the current driving the laser diode, and the position of a grating reflecting some laser light back into the diode, cancelling out high frequency noise. To compensate for long-term drifts, the length of the SHG cavity is stabilised to the Rubidium D$_1$ transition at $795$~nm via a mirror mounted on a piezo-electric transducer. After leaving the SHG cavity, the remaining red light is separated from the blue pump light by a dichroic mirror, with the cavity resonance being monitored at PD$_2$. The pump light is modulated at $12.175$~MHz (EOM$_3$) to prevent beating with the formerly applied $12.5$~MHz, and coupled into the downconversion cavity. 
 
The SPDC process takes place in a dual anti-reflective coated $25$~mm long ppKTP crystal (coating Layertec, crystal Raicol) with a poling period of $8.8\ \mu m$, cut for type-II quasi-phase-matching. The crystal is placed in a copper mount that is temperature controlled with a stability of $\Delta$T $< \pm1.5$~mK. The bow-tie cavity surrounding the crystal is designed following \cite{Boyd1968,Abitan2004} and mounted on a monolith of Invar to increase stability. The cavity consists of three high-reflecting mirrors plus one partially reflecting outcoupling/incoupling mirror at $795$~nm/$397.5$~nm (all Layertec, see supplementary material). The half-wave plate (HWP) inside the cavity rotates the polarisation of the single photons by $90$\degr every physical round-trip, but leaves the pump light unaffected. This ``flip-trick'' effectively doubles the cavity length for the single photons, reducing their free spectral range to $FSR_r \approx 121$~MHz compared with $FSR_b \approx 241$~MHz. It also cancels out any birefringence between the orthogonally polarised single photons, as on average the number of round-trips travelled in each polarization is the same. Thereby the flip-trick alleviates the need for a compensation crystal \cite{Wolfgramm2008,Scholz2009,Zhou2014}, reducing the overall loss and achieving finesses of ${\cal F}_r \approx 181$ and ${\cal F}_b \approx 8.5$.

In order to achieve enhanced emission into the desired mode while simultaneously maintaining a $100$\% duty cycle of the source, the cavity needs to be kept triply resonant. To ensure this, a complex active stabilisation circuit for signal, idler and pump photons has been implemented. The first step uses a signal derived from PD$_3$ to stabilise the length of the cavity to the pump frequency via a mirror mounted on a piezo-electric transducer. Secondly, the flip-trick always ensures double resonance of signal and idler, as it eliminates birefringence of the single photons. Temperature tuning the ppKTP crystal allows to fulfil the phase-matching condition for SPDC while simultaneously achieving triple resonance at temperatures of around T $\approx 41.3^{\circ}$~C with a $FWHM \approx 10$~mK. Conventionally, the $FSR$ of a cavity is independent of the frequency of the light field, resulting in twice the number of resonances for every second harmonic frequency (pump) compared with its fundamental (single photons). Locking to an unmatched resonance would yield no useful single photons. The flip-trick compensates for this as well by effectively doubling the cavity length only for the fundamental frequency, ensuring every resonance of the pump is matched with one for the single photons. 

After leaving the cavity, a dichroic mirror and an ultra-narrow bandpass filter at $795$~nm (FWHM $1.0$~nm) seperate the down-converted photons from the leaking pump light. A polarising beamsplitter deterministically splits the paired photons, which are then separately coupled into single mode fibres, detected on single photon detectors (Perkin Elmer SPCM-AQR-14-FC) and recorded with a time-tagging module (Roithner Lasertechnik, $100.1$~ps time resolution). The self-relocking optical stabilisation loops for the laser frequency, the SHG and the PDC cavity lengths commonly allow continuous data acquisition for $90$ minutes, and regularly reach up to half a day.
 
 
\section{Results}

The characterisation of the source is performed by measuring the temporal intensity cross-correlation function \Gt between signal and idler photon. The cross-correlation function is given by \cite{Herzog2008,Scholz2009}:

\begin{equation}\label{eqn:G2}
\begin{split}
G^{(2)}_{s,i}(\tau) &= \bra{\Psi}E_I^{(-)}(t)E_S^{(-)}(t+\tau)E_S^{(+)}(t+\tau)E_I^{(+)}(t)\ket{\Psi} \\
&\propto  \Bigg| \sum_{m_s,m_I} \frac{\sqrt{\gamma_S\gamma_I\omega_s\omega_I}}{\Gamma_s\Gamma_I} \\
&\quad \times 
\begin{cases}
e^{-2 \pi \Gamma_s (\tau-(\tau_0/2))}{\rm sinc}{(i \pi \tau_0 \Gamma_s)} & \forall \hspace{0.5 mm} \tau \geqslant \frac{\tau_0}{2}\\
e^{+2 \pi \Gamma_i (\tau-(\tau_0/2))}{\rm sinc}{(i \pi \tau_0 \Gamma_i)} & \forall \hspace{0.5 mm} \tau < \frac{\tau_0}{2}
\end{cases} \Bigg|^2,
\end{split}
\end{equation}

\noindent where, $\forall \hspace{0.5 mm} k \in \{S,I\}$,  $E_k^{(\pm)}$ are the electric field operators, $\gamma_{k}$ is the cavity damping rate, $\omega_{k}$ is the single photon frequency, $\Gamma_k = \frac{\gamma_k}{2}+i\,m_k\text{\textit{FSR}}_{r,k}, \hspace{0.5 mm} m_k \in \mathbb{Z}$ and $\text{\textit{FSR}}_{r,k}$ is the free spectral range. $\tau_0$ corresponds to the temporal width of the peaks, accounting for the propagation delay between signal and idler in the crystal.

\begin{figure}[!t]
\includegraphics[width=\linewidth]{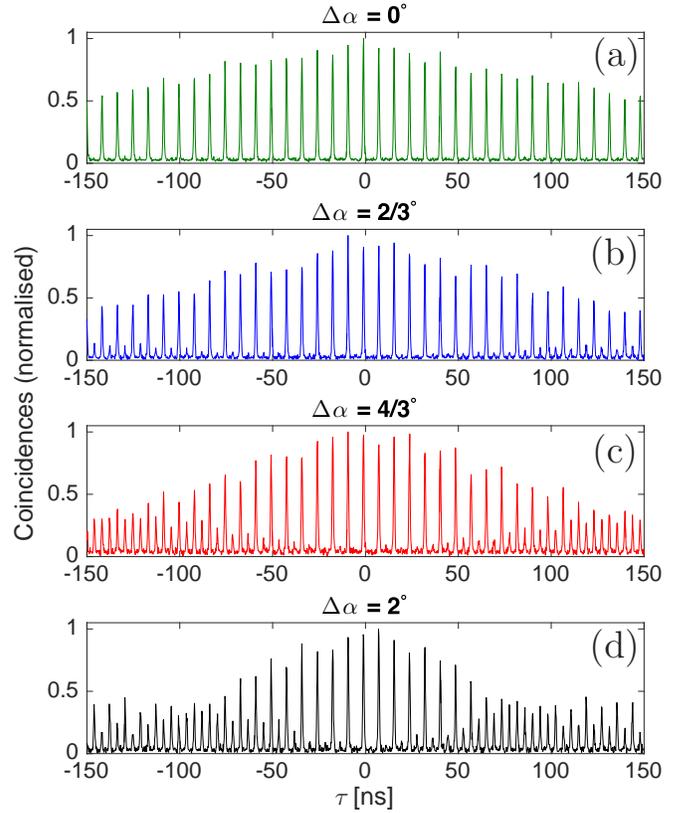}
\put(-27,282){\Large(a)}
\put(-28,207){\Large(b)}
\put(-27,132){\Large(c)}
\put(-28,58){\Large(d)}
\caption{(a)-(d) Evolution of the normalised temporal intensity cross-correlation function \Gt as the HWP is detuned from the optimum by (a) $\Delta\alpha =0\,$\degr, (b) $\sfrac{2}{3}\,$\degr, (c) $\sfrac{4}{3}\,$\degr\ and (d) $2$\degr. This detuning reveals an additional cavity with half the temporal spacing, with its maximum around (b) $\tau > \pm 150$~ns, (c)~$\tau \approx \pm 140$~ns and (d) $\tau \approx \pm 90$~ns.}
\label{fig:evolution}
\end{figure}

\begin{figure*}[!t]
  \setlength\abovecaptionskip{0cm}
  \centering
  \begin{minipage}[b]{0.45\textwidth}
    \begin{overpic}[height=5cm]{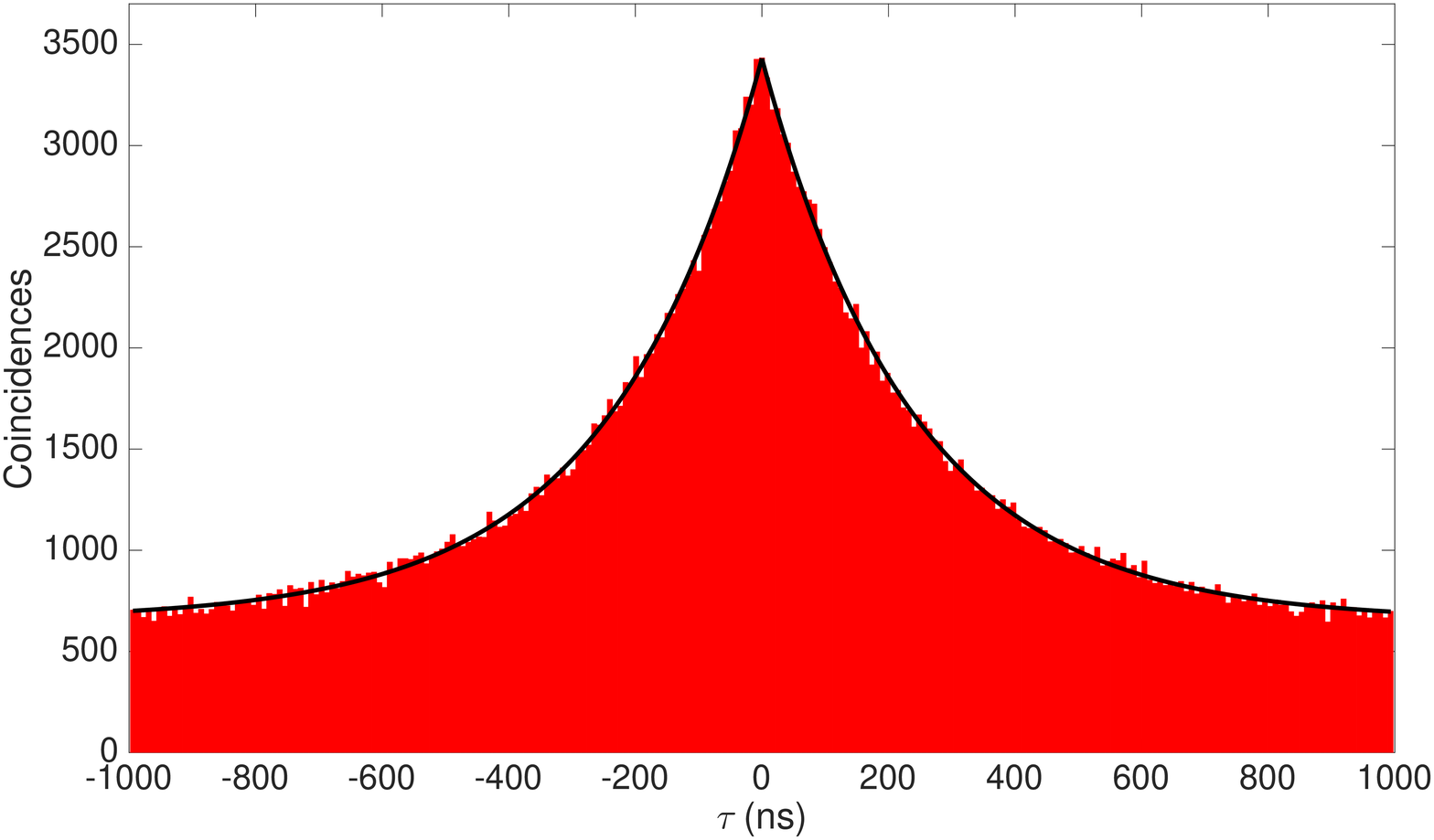}
      \hypersetup{hidelinks}
      \put(85,50){\Large(a)}
    \end{overpic}
  \end{minipage}
  \qquad
  \begin{minipage}[b]{0.45\textwidth}
    \begin{overpic}[height=5cm]{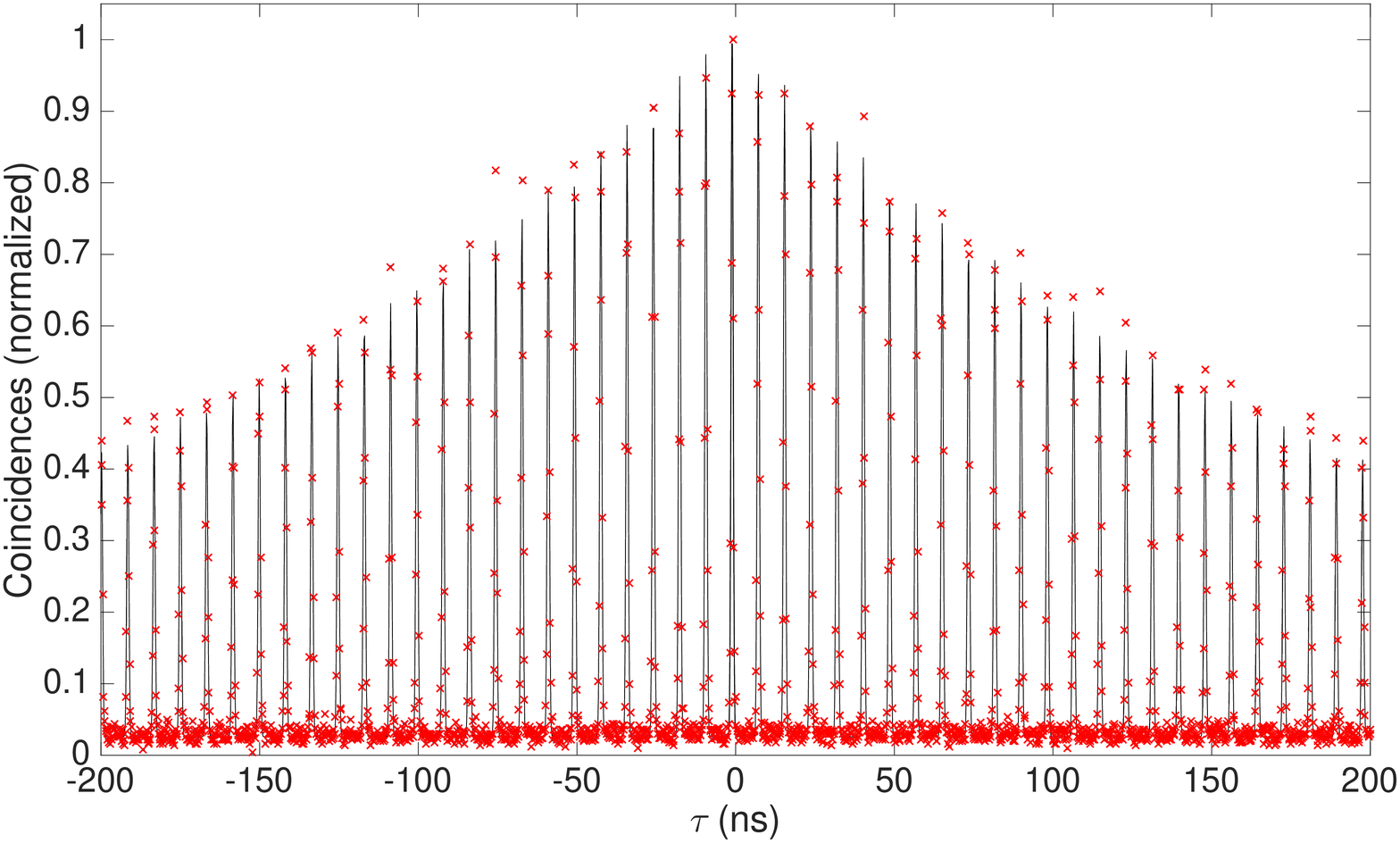}
      \hypersetup{hidelinks}
      \put(85,50){\Large(b)}
    \end{overpic} 
  \end{minipage}
  \caption{Temporal intensity cross-correlation function \Gt of the source for perfect HWP alignment. Data in red, theoretical model in black. (a) Time bin size of $8.2$~ns chosen to be close to one effective round-trip time. Linewidths of signal and idler are derived from the double exponential decay to be $666 \pm 16$~and $667 \pm 15$~kHz. Overall counts are $312,000$ within a $\pm 1\ \mu$s window and an integration time of 11~min. (b) Zoom into the detailed structure with the peaks normalised to the main peak at zero delay. Time bin size is 200.2~ps. Average pump power over the measurement is $20\ \mu$W with fluctuations $< 5\%$.}
   \label{fig:coinc}
\end{figure*}

Fig.~\ref{fig:evolution} shows measurements of \Gt for four different angular configurations of the HWP without background correction. The comb-like structure arises from the increased probability of detecting signal and idler photons at integer multiples of their effective round-trip time, $t_{rt} = 1/\text{\textit{FSR}}_{r,k} = 8.28$~ns, equaling two physical round-trips. At perfect alignment (Fig.~\ref{fig:evolution}\hyperref[fig:evolution]{a}), the HWP rotates horizontal to vertical polarisation and vice versa at each pass. Therefore, the photon mode inside the cavity does not overlap with itself one physical round-trip earlier. It has to traverse the cavity a second time to complete an effective round-trip and get detected with its partner as a coincidence on the SPDs. This halves the FSR and thus reduces the photon bandwidth. Additionally, it ensures double resonance of signal and idler photons by cancelling the birefringence of the ppKTP crystal. Slight misalignment of the HWP angle results in a non-zero probability for the photons to be detected as coincidences after an odd number of physical round-trip differences. This can be seen in the cross-correlation functions (Fig.~\ref{fig:evolution}) as peaks with half the temporal spacing, which reach their maximum when the round-trip difference between the signal and idler photons is an integer multiple of $\frac{45^{\circ}}{\Delta\alpha}$. Thus changing $\Delta\alpha$ from $0$\degr\ to $45$\degr\ (optical axis of HWP) allows one to set  $FSR_{r,0} \approx 121$~MHz and $FSR_{r,45} \approx 242$~MHz.

In case of perfect alignment of the HWP, fitting a double exponential decay $\text{exp}(-2\pi\Delta\nu\tau)$ can be used to extract the bandwidth $\Delta\nu$ of the single photons, as shown in Fig.~\ref{fig:coinc}\hyperref[fig:coinc]{a}). The FWHM correlation time of $331$~ns, and the corresponding bandwidths of $\Delta\nu_S = 666 \pm 16$~kHz and $\Delta\nu_I = 667 \pm 15$~kHz, are, to our knowledge, the best values observed from SPDC to date. The source therefore meets the spectral requirements for efficient coupling with the D$_1$ transition in Rubidium in GEM, making the photons suitable carries for quantum information between memories. The slight difference between signal and idler bandwidth is within error of the fit. Fig.~\ref{fig:coinc}\hyperref[fig:coinc]{b}) shows a zoom into delay times of $\pm 200\ \mu\text{s}$ around zero delay with a finer resolution of $200.2$~ps per time bin. The fit of Eqn.~\ref{eqn:G2} is in excellent agreement with the data. The finesse of ${\cal F}_r \approx 181$ corresponds to internal losses of $3.4$\% per effective round-trip, leading to an escape efficiency of $\sim0.29$ \cite{Wolfgramm2011a}.


\section{Conclusion and Outlook}

We report on the first sub-MHz cavity-enhanced single photon source, resonant with the D$_1$ transition in Rubidium. The linewidth of \myhref{https://www.youtube.com/watch?v=WxnN05vOuSM}{$666$}~kHz is the narrowest from SPDC to date, matching the spectral requirements of GEM---currently the most efficient quantum memory. The ultra-narrow bandwidth is achieved by introducing a new method, the flip-trick, which also cancels out birefringence between orthogonally polarised single photons. Average locking times of several hours are achieved and stabilising the cavity to the pump light enables a $100$\% duty cycle. The next step will be to filter a single cavity mode by using either active spectral filtering in a cavity, the clustering effect \cite{Pomarico2012,Fekete2013}, or an etalon. Additionally, the whole setup fits on two small breadboards, which are easy to transport, and we are planning on combining our source with the GEM developed at the Australian National University to gain further insight into the memory's performance at a single photon level.


\section*{Funding Information}
This work was partially supported by the Centre for Quantum Computation and Communication Technology (Grant No. CE110001027), and by the Centre for Engineered Quantum Systems (Grant No. CE110001013). AGW acknowledges support from a UQ Vice-Chancellor's Research and Teaching Fellowship.


\section*{Acknowledgments}

The authors thank the team from the Austrian Institute of Technology and Roithner LaserTechnik for kindly providing time-tagging modules.



\bibliography{biblio_rb_project_arxiv}

\clearpage

\section{Supplementary Material}

\subsection{PDC cavity specification}

The parametric down conversion (PDC) process takes place in a bow-tie cavity with two curved mirrors. The cavity length is $\sim124$~cm, the distance between the curved mirrors being $\sim21.5$~cm. Coating specifications at both $397.5$~nm and $795$~nm are listed in Table~\ref{tab:pdc}.

\begin{table}[htbp]
  \centering
  \caption{PDC cavity mirror reflectivities at $397.5$~nm and $795$~nm. M1 and M4 are plane, M2 and M3 are plano-concave with radius of curvature of $-200$~mm.}
  \label{tab:pdc}
  \begin{tabular*}{\linewidth}{@{\extracolsep{\fill}}lcc}
    \hline
    &Coating at $397.5$~nm &Coating at $795$~nm\tabularnewline\hline
    M1 (Incoupling) &$\mathrm{PR} = 98.0\%\pm0.4\%$ &$\mathrm{HR} > 99.9\%$ \tabularnewline
    M2, M3 &$\mathrm{HR} > 99.85\%$ &$\mathrm{HR} > 99.9\%$ \tabularnewline
    M4 (Outcoupling) &$\mathrm{HR} > 99.85\%$ &$\mathrm{PR} = 99.0\%\pm0.2\%$ \tabularnewline\hline
  \end{tabular*}
\end{table}

\subsection{Cavities' finesses and free spectral ranges including uncertainties}

The cavity in which light at $397.5$~nm is generated via second harmonic generation (SHG) is another bow-tie with two curved mirrors. The outcoupling mirror is anti-reflective coated at that wavelength to make the cavity ``invisible'' for blue light. The finesse and free spectral range of the SHG cavity at $795$~nm, as well as of the PDC cavity for both red and blue, are given in Table~\ref{tab:fsr-finesse}.

\begin{table}[htbp]
  \centering
  \caption{Measurements of finesse and free spectral range (FSR) for both the PDC and SHG cavities.}
  \label{tab:fsr-finesse}
  \begin{tabular*}{\linewidth}{@{\extracolsep{\fill}}lcu{5}u{3}}
    \hline
    & Wavelentgh &\multicolumn{1}{c}{FSR} &\multicolumn{1}{c}{Finesse, $\cal F$}\tabularnewline\hline
    SHG &$795$~nm &278!3\text{ MHz} &100!2\tabularnewline
    PDC &$795$~nm &120.8!0.1\text{ MHz} &181!4\tabularnewline
    PDC &$397.5$~nm &241!2\text{ MHz} &8.5!0.3\tabularnewline\hline
  \end{tabular*}
\end{table}

\end{document}